\documentclass[aps,prd,nofootinbib,twocolumn,preprintnumbers]{revtex4}
\usepackage{graphicx,amsmath,amssymb}
\newcommand{\bm}[1]{\mbox{\boldmath $#1$}}
\usepackage{ulem}
\usepackage[usenames]{color}
\definecolor{DarkGreen}{rgb}{0.0,0.5,0.0}

\begin{document}
\title{Cosmological constraints on rapid roll inflation}
\author{Takeshi Kobayashi}\email{takeshi.kobayashi@ipmu.jp}
\affiliation{{Institute for Cosmic Ray Research, The University of
Tokyo, 5-1-5 Kashiwanoha, Kashiwa, Chiba 277-8582, Japan}}
\author{Shinji Mukohyama}\email{shinji.mukohyama@ipmu.jp}
\author{Brian A. Powell}\email{brian.powell@ipmu.jp}
\affiliation{Institute for the Physics and Mathematics of the Universe (IPMU), The
University of Tokyo, 5-1-5 Kashiwanoha, Kashiwa, Chiba 277-8582, Japan}
\date{\today}
\preprint{IPMU09-0050}
\preprint{ICRR-Report-543}
\begin{abstract}
\noindent
We obtain cosmological constraints on models of inflation which exhibit rapid roll
solutions.  Rapid roll attractors exist for potentials with large mass terms and
are thus of interest for inflationary model building within string theory.  We
constrain a general ansatz for the power spectrum arising from rapid roll
inflation that, in the small field limit,  can be associated with tree level hybrid potentials with variable mass terms
and nonminimal gravitational coupling $\xi R\phi^2$.  We consider perturbations generated through
modulated reheating and/or curvaton mechanisms in place of the observationally
unacceptable primary spectra generated by inflaton fluctuations in these models.
The lack of a hierarchy amongst higher-order $k$-dependencies of the power spectrum
results in models with 
potentially large runnings, allowing us to impose tight constraints on such models using
CMB and LSS data. In particular, we find $n_s <1$ and $|\alpha| < 0.01$.  We conclude with a concrete realization of rapid roll inflation
within warped throat brane inflation that is in good agreement with current data. 
\end{abstract}
\maketitle
\section{Introduction}

Recent years have seen steady progress in the development of inflation within string
theory \cite{Cline:2006hu,Kallosh:2007ig,Burgess:2007pz,McAllister:2007bg,Baumann:2009ni}.  
Inflation in string theory is beset by the $\eta$-problem
\cite{Copeland:1994vg,Kachru:2003sx}, in which 
dimension 6, Planck-suppressed operators induce corrections to the inflaton mass
of order the Hubble scale, $\delta m \sim \mathcal{O}(H)$.
Such models are characterized by the parameter $\eta = M^2_{\rm Pl}\frac{V''}{V} \sim
\mathcal{O}(1)$, preventing slow roll inflation.  These corrections generically arise in the volume stabilization of brane
inflation \cite{Kachru:2003sx}, and a compelling solution to this problem remains an active
area of research.  This problem is compounded in models based on warped throat
compactifications in which the length of the throat restricts the maximal field
range \cite{Baumann:2006cd}.  It was earlier realized by Linde that it is
possible to achieve significant efoldings of expansion even in the presence of large mass
terms \cite{Linde:2001ae} if the field enters a stage of {\it rapid roll} inflation.  This choice of
terminology will become clear below.

In this paper, we obtain observational constraints on models of inflation that
undergo a period of rapid roll inflation.  Typical examples of such potentials are
those appearing in models of tree-level hybrid inflation, in which the field evolves
along the rapid roll attractor, asymptotically approaching the
minimum. These 
potentials give blue scalar spectra, which is observationally 
disfavored unless the tensor contribution is significant
\cite{Komatsu:2008hk}, and so we
consider perturbations generated through other mechanisms, such as modulated
reheating or the curvaton scenario.\footnote{Another good example of
rapid roll inflation is hilltop inflation~\cite{Boubekeur:2005zm}
taking place near a maximum of the potential with $\eta \sim
-\mathcal{O}(1)$. In this case the spectral tilt of perturbations
generated by the inflaton becomes (highly) red.} The light fields
necessary for generating the primordial perturbations in these scenarios
are abundant in string theory, making such mechanisms natural in this
setting \cite{Kobayashi:2009cm}. 

We derive an analytical expression for the power spectrum, $P(k)$, arising from potentials of
the form
\begin{equation}
\label{v}
V = V_0\left[ 1 + \frac{1}{2}v_2\frac{\phi^2}{M_{\rm Pl}^2}
	+ \mathcal{O}\left(\frac{\phi^4}{M_{\rm Pl}^4}\right)\right]
\end{equation}
in which the field is restricted to the range $\phi \ll M_{\rm Pl}$.  The formalism we
develop is applicable to both minimally and nonminimally coupled scalar fields.  The
key result is that the power spectrum arising from modulated reheating and/or
curvatons does not possess a hierarchy amongst low-order
and high-order $k$-dependencies, in contrast to inflaton-generated perturbations.
  In fact, the higher-order runnings of the spectrum are completely determined by the
spectral index, $n_s$, and first-order running, $\alpha =dn_s/d{\rm ln}k$, in the regime where
Eq. (\ref{v}) is justified.  Using the latest
CMB and LSS data, we perform a Bayesian analysis on this power spectrum.  We compare these constraints with those obtained on a
general ansatz for the power spectrum, $P(k) \propto k^{(n_s - 1) + \frac{1}{2}\alpha
{\rm ln}\frac{k}{k_0}}$, where the parameters $n_s$ and $\alpha$ are allowed to vary
freely.  The spectrum arising from rapid roll inflation is highly constrained
relative to the generic parameterization, favoring $n_s <1 $ and small running,
$|\alpha| < 0.01$.  

Lastly, we present a concrete realization of rapid roll inflation within the context
of warped throat brane inflation.  We find that it is possible to obtain sufficient
inflation in the presence of a range of mass terms, and the predictions of
the model are in good agreement with the observational constraints we obtain in this
paper. 

In Section II, we review the concept of rapid roll inflation.  In section III, we
discuss the generation of perturbations in these models through either modulated
reheating or curvatons, and derive an analytic expression for $P(k)$.  In section IV
we present the details of the Bayesian analysis, and report our constraints on the
power spectrum.  In Section V, we present a specific realization of rapid roll
inflation within string theory.

\section{Rapid Roll Inflation}

In this section we review the phenomenon of rapid roll inflation. To
illustrate this concept, consider a minimally coupled scalar field
$\phi$ with a potential of the form 
\begin{equation}
\label{pot}
V(\phi) = V_0 + \frac{1}{2}m^2\phi^2.
\end{equation}
This is a prototypical hybrid-type potential, in which inflation ends
when an auxiliary field develops a tachyonic instability around 
$\phi\approx 0$. The equation of motion of a homogeneous scalar field in
an FRW universe with this potential is 
\begin{equation}
\ddot{\phi} + 3H\dot{\phi} + m^2\phi = 0.
\end{equation}
When $H \approx $ const. we obtain the solution
\begin{eqnarray}
\label{sol}
\phi(t) &=& \phi_+ e^{-r_+Ht}  + \phi_- e^{-r_- Ht} \\
r_{\pm} &=& \frac{3}{2} \mp\sqrt{\frac{9}{4} - \frac{m^2}{H^2}},
\end{eqnarray}
where $\frac{m^2}{H^2} < \frac{9}{4}$.  This is true for the potential
Eq. (\ref{pot}) when $r_+^2\phi^2\ll M_{\rm Pl}^2$ and 
$m^2\phi^2\ll H^2M_{\rm Pl}^2$. The general solution is a superposition
of both branches of Eq. (\ref{sol}) but at late times the field evolves
along the $r_+$ branch, which is characterized by 
\begin{equation}
\label{rr}
 \dot{\phi} \simeq -r_+ H\phi.
\end{equation}
Note that this behavior exists even if $m/H=O(1)$, i.e. one of the slow
roll conditions is violated as $\eta=O(1)$. When $m/H=O(1)$, the
dynamics defined by the solution Eq. (\ref{rr}) is termed {\it rapid 
roll} inflation. It tells us that the logarithmic variation of $\phi$ is set by the
Hubble scale, or $\Delta\phi=O(\phi)$ in a Hubble time 
$\Delta t=H^{-1}$. This is in contrast to {\it slow roll} inflation, for 
which $|\dot{\phi}| \ll |H\phi|$. The number of efolds of expansion
obtained during rapid roll is 
\begin{equation}
N = \int H dt \simeq -\frac{1}{r_+}\int \frac{d\phi}{\phi} = 
 \frac{1}{r_+}{\rm ln}{\frac{\phi_i}{\phi_f}},
\end{equation}
where $\phi_i$ and $\phi_f$ are the field values at the beginning and
end of inflation. The spectral index of curvature perturbations from the
inflaton is given by  $n_s-1=2r_+$, and is therefore blue for this class
of models \cite{GarciaBellido:1996ke}.\footnote{The horizon crossing
formalism breaks down here, and care must be taken when evaluating the
spectrum \cite{Kinney:2005vj,Tzirakis:2007bf}.} When $m/H=O(1)$, this
means that we need other sources of curvature perturbations, such as
curvaton and/or modulated reheating, to confront with observational
data. (See the beginning of the next section for more about this point.)

Rapid roll solutions exist if a potential is approximated by (\ref{pot})
and if the field value is small enough. In order to make this statement
quantitative, we consider more general potentials and introduce
parameters to measure deviation from the form (\ref{pot}) and smallness
of a field value. As shown in the appendix, the inflationary dynamics
can be described by
\begin{equation}
\label{gfr}
cH\dot{\phi}(t) \simeq -V'(\phi)
\end{equation}
with $3M_{\rm Pl}^2H^2\simeq V$ if $\epsilon \ll 1$ and 
$|\bar{\eta}| \ll 1$ are satisfied. Here 
\begin{eqnarray}
\label{Eps}
\epsilon &=& \frac{M_{\rm Pl}^2}{2}\left(\frac{V'}{V}\right)^2,\\
\bar{\eta} &=& \eta + \frac{c^2}{3} - c,\\
\label{Eta}
\eta &=& M_{\rm Pl}^2\frac{V''}{V} 
\end{eqnarray}
where $c$ is a constant of order unity. The value of $c$ is determined
so that $|\bar{\eta}|$ is minimized in the duration of interest and that
the corresponding solution is a dynamical attractor (see the appendix
for details). For example, for the proto-type potential (\ref{pot}) with
$\phi^2/M_{\rm Pl}^2$ small enough, there are two values of $c$
minimizing $|\bar{\eta}|$. The larger value corresponds to an attractor
and agrees with the $r_+$ branch (\ref{rr}). Roughly speaking,
$\bar{\eta}$ measures deviation of $V$ from the form (\ref{pot}) and
$\epsilon$ measures smallness of the field.

Rapid roll inflation can also be manifested in models of inflation
driven by non-minimally coupled scalar fields. In the rest of this
section, for simplicity we consider a conformally coupled scalar field
to drive rapid roll inflation. More systematic analysis with generic 
couplings will be presented in the next section. Thus, in the following
we shall first write down some formulas with general couplings for later
convenience and then restrict our consideration to the conformal
coupling.

Consider the action of a non-minimally coupled scalar, 
\begin{equation}
\label{ac1}
S = \int d^4x\sqrt{-g}\left[\frac{R}{2M_{\rm Pl}^2} - \frac{1}{2}\partial^\mu\phi
\partial_\mu \phi - V(\phi) - \frac{1}{2}\xi R \phi^2\right].
\end{equation}
In an FRW universe with $ds^2 = -dt^2 + a(t)^2d{\bf x}^2$, the resulting
equations of motion are
\begin{eqnarray}
\label{ce1}
3M_{\rm {Pl}}^2 H^2 &=& \frac{1}{2}\dot{\phi}^2 + 6\xi H \phi \dot{\phi} + 3\xi \phi^2 H^2
+
V(\phi), \\
\label{ce2}
 M_{\rm Pl}^2\dot{H}& = &
  -\frac{1}{2}(1-2\xi)\dot{\phi}^2 - 4\xi H\phi\dot{\phi} -12\xi^2H^2\phi^2 \nonumber \\
&-& \xi \phi V'(\phi) + \xi(1-6\xi)\phi^2\dot{H}, \\
\label{ce3}
\ddot{\phi} + 3H\dot{\phi} &+& 6\xi (\dot{H} + 2H^2)\phi + V'(\phi) = 0.
\end{eqnarray}
The value $\xi = \frac{1}{6}$ is known as conformal coupling.  In this
limit, the non-potential terms on the right hand side of Eq. (\ref{ce1})
can be written as a perfect square, 
\begin{equation}
3M^2_{\rm Pl} H^2 = (\dot{\phi} + H\phi)^2 + V(\phi).
\end{equation}
Then, for constant potential, $V(\phi) = V_0$, the rapid roll solution
$\dot{\phi} = -H\phi$ gives a de Sitter universe,
\begin{equation}
\label{dS}
3M^2_{\rm Pl} H^2 = V_0.
\end{equation}
The existence of rapid roll solutions in this model could be understood
by recalling that for $\dot{H} \approx 0$, the Ricci scalar 
$R \approx 12H^2$.  Then for $\xi = \frac{1}{6}$, the coupling term in
Eq. (\ref{ac1}) forms an effective mass term $\simeq H^2\phi^2$, and we 
recover the potential Eq. (\ref{pot}). On the other hand, the exact de
Sitter symmetry (\ref{dS}) in the conformally coupled frame could not
have been guessed from the analysis for the minimally coupled models
presented in the previous paragraphs. This difference between the
conformal model with a flat potential and the minimal model with
$m^2\simeq 2H^2$ becomes important when we discuss observables such as
the spectral tilt of curvature perturbations generated from curvatons
and/or modulated reheating since they reflect tiny deviations from the
exact de Sitter symmetry. The conformally-coupled rapid-roll inflation
was first investigated in \cite{Kofman:2007tr}, where it was shown that
rapid roll attractor solutions exist if a potential is flat enough. A
small deviation of the non-minimal coupling strength $\xi$ from $1/6$
was considered in \cite{Kobayashi:2008rx}. In the concrete example
considered in \cite{Kofman:2007tr}, the conformal coupling arises as the
Hubble scale mass correction from volume stabilization in the KKLMMT
\cite{Kachru:2003sx} model and the potential is of the form 
\begin{equation}
V(\phi) = V_0\left[1 - \left(\frac{M_{\rm Pl} \Delta}{\phi}\right)^4\right],
\end{equation}
where $\Delta$ is a dimensionless constant.  This potential is flat
except for very small values of $\phi$, and so the de Sitter solution
holds across much of the inflationary trajectory.

In this paper, we seek constraints on rapid-roll inflation with not only
minimal and conformal couplings but also arbitrary non-minimal coupling
strengths.

\section{Primordial Perturbations}

In this section we derive an analytic expression for the power spectrum generated by the
rapid roll models introduced in the last section.  An important aspect of this work is that
we consider perturbations {\it not} generated by the inflaton itself, but rather from a mechanism
such as modulated reheating \cite{Dvali:2003em,Kofman:2003nx} or curvatons
\cite{Lyth:2001nq,Moroi:2001ct,Lyth:2002my}.  We do this for several reasons. 
The first is that spectra arising from the minimally coupled inflaton 
with potential Eq. (\ref{pot}) or flat potentials with almost conformal 
couplings $\xi \approx 1/6$ are blue ($n_s > 1$) with small tensors. 
Observationally, blue spectra with low tensor-to-scalar ratio are 
disfavored \cite{Komatsu:2008hk}.
In addition to phenomenological motivations, such mechanisms are also attractive from a theoretical
point of view.  In particular, string theory provides natural candidates for the light
moduli fields that generate perturbations in these alternative scenarios.  
Current work in which the curvaton is identified with an open string mode
is under investigation \cite{kobayashi}.
The low energy scales required to suppress the inflaton perturbations on CMB scales are
well tolerated by string inflation models, particularly brane inflation in warped throat
compactifications.  
String inflation is
also a pertinent place to realize the rapid roll inflation models considered in this paper.

\subsection{Modulated Reheating and Curvatons}

In order to reheat the universe, the inflaton field must undergo decays whose end products
are the particles of the Standard Model.  In supersymmetric and string theories, the
couplings that facilitate these decays are not constants, but functions of scalar fields in
the theory.  If these fields are light during inflation ($m \ll H$), they will fluctuate in
space with the result that the coupling constants (and hence decay rates) will become
functions of space.  Different parts of the universe thus decay and reheat at different
times -- the resulting spatial modulations in the reheat temperature give rise
to energy density perturbations,
\begin{equation}
\frac{\delta T_{\rm RH}}{T_{\rm RH}} \propto \frac{\delta \rho}{\rho}.
\end{equation}
For a light modulus field $\chi$, one finds \cite{Dvali:2003em}
\begin{equation}
\frac{\delta \rho}{\rho} \propto \frac{\delta \chi}{M},
\end{equation}
where $M$ is some mass scale.  The key result is that the power spectrum of the resulting
fluctuations is then $\propto H^2/M^2$, where $M$ can be taken sufficiently smaller than
$M_{\rm Pl}$ to ensure that the inflaton perturbations are subdominant. 

The primordial density perturbation may also be generated via the
curvaton mechanism
\cite{Lyth:2001nq,Moroi:2001ct,Lyth:2002my}.  The curvaton, $\sigma$, is
a light scalar during inflation with an associated vacuum fluctuation,
$\delta \sigma$.  These fluctuations constitute a superhorizon
isocurvature perturbation that grows during a post-inflationary radiation
dominated phase.  After curvaton decay, the perturbation is converted
into an adiabatic mode with amplitude \cite{Lyth:2001nq}
\begin{equation}
P_{\delta} \propto \frac{1}{\pi^2}\left(\frac{H}{\sigma}\right)^2.
\end{equation}
If the curvaton is sufficiently light during inflation, $\sigma = const$
as a result of the Hubble drag and $P_\delta \sim H^2$.  As in the case
of modulated reheating by decaying light particles, the power spectrum
is completely determined by the evolution of the Hubble parameter.

\subsection{Power Spectrum}

We now consider a rapid roll inflation driven by a non-minimally coupled
inflaton $\phi$ and obtain an analytic expression for the power spectrum
of curvature perturbations generated from curvaton/modulated
reheating. Note that we consider minimally coupled field(s) responsible
for curvatons/modulated reheating. This is possible if those fields
responsible for curvature perturbations are associated with shift
symmetries. For example, in the stringy setup presented
in~\cite{Kobayashi:2009cm}, minimally coupled, light scalars can arise
from angular isometries of the warped deformed conifold. Those light
fields are responsible for curvaton/modulated reheating. On the other
hand, we consider a non-minimally coupled scalar field as an
inflaton. (Of course this class of models includes a minimally coupled
inflaton as a special case.) As we shall see explicitly, observables
such as the spectral tilt and its scale-dependence depend on the
non-minimal coupling strength in a way that cannot be mimicked by a 
potential. Indeed, in the regime of validity of an approximation, we
shall see that the scale-dependence of the power spectrum is
characterized by two independent parameters, one of which is the
non-minimal coupling strength and the other of which is a parameter of
the potential. This explicitly shows that the effect of non-minimal
coupling to observables cannot be mimicked by potentials in the regime
of validity of the approximation.

Note again that we shall consider curvature perturbations generated from
curvaton/modulated reheating. However, the background FRW evolutions is
of course determined by the dynamics of the inflaton $\phi$, which is in
general non-minimally coupled. Therefore, we consider a non-minimally
coupled scalar field with a potential of the form 
%
\begin{eqnarray}
 V &=& V_0
  \left[1 + 
   \sum_{n=1}^{\infty}\frac{v_{2n}}{(2n)!}x^n\right], \\
x &\equiv&  \frac{\phi^2}{M_{\rm Pl}^2}, 
\end{eqnarray}
where $V_0$ and $v_{2n}$ are constants, and we suppose that $x \ll 1$. 
This is nothing but the Taylor expansion of a general potential with the
$Z_2$ symmetry $\phi\to -\phi$, and can be considered as a simple
generalization of (\ref{pot}). In the following, we shall perform a
systematic analysis of the system by expansion w.r.t. $x$ ($\ll 1$). 
The scalar field satisfies the equations of motion Eqs. (\ref{ce1})-(\ref{ce3}).
Let us define $\Delta_1$ and $\Delta_2$ by
%
\begin{eqnarray}
 \Delta_1 & \equiv & \frac{3M_{\rm Pl}^2H^2}{V_0}
  - \left[1 + \sum_{n=1}^{N+1}
     \alpha_{2n}x^n \right], \nonumber\\
 \Delta_2 & \equiv & \frac{\dot{\phi}}{H\phi} + \sum_{n=0}^N\beta_{2n}x^n,
\end{eqnarray}
where $N$ is non-negative integer, and $\alpha_{2n}$ and $\beta_{2n}$
are constants to be determined. Note that the smallness of the
parameters $\Delta_1$ and $\Delta_2$ guarantees the inflaton to be on
the rapid roll attractor. (For the case of a minimally-coupled inflaton,
each corresponds to approximations (\ref{eq-3}) and (\ref{eq-4}).) 
We shall determine $\alpha_{2n}$ with
($n=1,\cdots,N+1$) by demanding that 
%
\begin{equation}
 \Delta_1 = O(x^{N+2}).
  \label{eqn:eq-Delta1}
\end{equation}
On the other hand, we demand that $\Delta_2$ satisfies 
%
\begin{equation}
 \frac{\dot{\Delta}_2}{H} + \gamma \Delta_2 = O(x^{N+1})
  \label{eqn:eq-Delta2}
\end{equation}
for a positive constant $\gamma$. This condition uniquely determines
$\beta_{2n}$ ($n=1,\cdots,N$) and $\gamma$ ($>0$). Note that the
positivity of $\gamma$ is required by the stability of
Eq. (\ref{eqn:eq-Delta2}): $\Delta_2\to O(x^{N+1})$  as $a\to\infty$.

We now set $N=0$ to obtain the lowest order result. In principle, we can
perform expansion up to larger $N$. For our purpose in the present
paper, however, $N=0$ suffices. We follow the three steps to complete
the program: 
First,
we eliminate $\dot{H}$, $H^2$ and $\ddot{\phi}$ in
Eqs. (\ref{eqn:eq-Delta1}) and (\ref{eqn:eq-Delta2}) by using 
Eqs. (\ref{ce2}), (\ref{ce1}) and (\ref{ce3}),
respectively. 
Next,
we determine $\gamma$ by demanding that the coefficient of $\dot{\phi}$
in Eq. (\ref{eqn:eq-Delta2}) after the first step should vanish. 
Lastly,
we expand Eqs. (\ref{eqn:eq-Delta1}) and (\ref{eqn:eq-Delta2}) with respect to $x$ and
solve them order by order. 

The result is
%
\begin{eqnarray}
\alpha_2 & = & \left(\frac{1}{2}-2\xi\right)\beta_0 - \xi,\nonumber\\
 \beta_0 & = & \frac{3\pm\sqrt{9-12(v_2+4\xi)}}{2}, \nonumber\\
 \gamma & = &3-2\beta_0.
\end{eqnarray}
Since $\gamma$ is required to be positive, we need to choose the ``$-$''
sign for $\beta_0$. Thus, 
%
\begin{eqnarray}
 \alpha_2 & = & \left(\frac{1}{4}-\xi\right)
  \left[3-\sqrt{9-12(v_2+4\xi)}\right] - \xi, \\
\label{ab}
 \beta_0 & = & \frac{3-\sqrt{9-12(v_2+4\xi)}}{2},\\
 \gamma & = & \sqrt{9-12(v_2+4\xi)}.
\end{eqnarray}
The attractor equation Eq. (\ref{eqn:eq-Delta2}) with $N=0$ and $\gamma>0$
implies that $\Delta_2\to O(\epsilon)$ as $a\to\infty$. Thus, we have
$\dot{\phi}+\beta_0H\phi\simeq 0$. This gives
%
\begin{equation}
 \phi \simeq \phi_0\left(\frac{a}{a_0}\right)^{-\beta_0},
\end{equation}
where $\phi_0$ and $a_0$ are the values of $\phi$ and $a$ at $t=t_0$. 

Then,  Eq. (\ref{eqn:eq-Delta1}) with $N=0$ implies that
%
\begin{eqnarray}
 P(k) \propto \left.\frac{3M_{\rm Pl}^2H^2}{V_0}\right|_{k=aH}
  &\simeq&
  1+\alpha_2\frac{\phi_0^2}{M_{\rm Pl}^2}
  \left(\frac{k^2}{a_0^2H^2}\right)^{-\beta_0}\nonumber \\
  &\simeq& 
  1+\alpha_2\frac{\phi_0^2}{M_{\rm Pl}^2}
  \left(\frac{3M_{\rm Pl}^2k^2}{a_0^2V_0}\right)^{-\beta_0}\nonumber\\
\end{eqnarray}
By choosing $a_0$ to be the scale factor at the pivot scale $k=k_0$, we
obtain  
%
\begin{equation}
 P(k) \simeq \frac{P(k_0)}{1+\alpha_2\frac{\phi_0^2}{M_{\rm Pl}^2}}
  \left[
  1+\alpha_2\frac{\phi_0^2}{M_{\rm Pl}^2}
  \left(\frac{k}{k_0}\right)^{-2\beta_0}\right],
\end{equation}
where $\phi_0$ is the field value at the pivot scale.  
This can be written more concisely in the form
\begin{equation}
\label{spec}
P(k) = \frac{P(k_0)}{1+A}\left[1 +
A\left(\frac{k}{k_0}\right)^{-B}\right],
\end{equation}
with
%
\begin{eqnarray}
 A & = & \alpha_2\frac{\phi_0^2}{M_{\rm Pl}^2}\nonumber\\
  &=& \left\{\left(\frac{1}{4}-\xi\right)
     \left[3-\sqrt{9-12(v_2+4\xi)}\right] - \xi\right\}
  \frac{\phi_0^2}{M_{\rm Pl}^2}, \nonumber\\ 
\\
 B & = & 2\beta_0 = 3-\sqrt{9-12(v_2+4\xi)}. 
\end{eqnarray}
This is the main result of this section. Note that the scale-dependence
of the power spectrum is characterized by the two parameters $A$ and
$B$, which are expressed in terms of $\xi$ and $v_2$. It is easy to see
that the effect of $\xi$ on ($A$, $B$) cannot be mimicked by $v_2$, and
vice versa.

A unique aspect of power spectra arising from rapid roll inflation is
the lack of a hierarchy amongst higher-order time derivatives of the
Hubble parameter and higher-order $k$-dependencies of the power
spectrum.  Such hierarchies do exist in slow roll inflation, with the
result that higher-order runnings are suppressed.

For  example, in the case of nearly conformal inflation with $\xi = 1/6
+ \delta \xi$ on a flat potential ($v_{2n} = 0$), we have 
\begin{eqnarray}
A &=& - [\delta \xi + \mathcal{O}(\delta \xi^2)]\frac{\phi_0^2}{M_{\rm Pl}^2},\\
B &=& 2(1+12\delta \xi) + \mathcal{O}(\delta \xi^2).
\end{eqnarray}
Then, the spectral index and runnings of the power spectrum Eq. (\ref{spec}) are
\begin{eqnarray}
\label{runnings}
n_s - 1 &\simeq& 2\delta \xi \frac{\phi_0^2}{M_{\rm Pl}^2},\\
\left.\left(\frac{d}{d{\rm ln}k}\right)^2{\rm ln}P(k)\right|_{k=k_0} &\simeq& -4\delta
\xi\frac{\phi_0^2}{M_{\rm Pl}^2},\\
&\vdots& \nonumber \\
\left.\left(\frac{d}{d{\rm ln}k}\right)^m{\rm ln}P(k)\right|_{k=k_0} &\simeq& -(-2)^m\delta
\xi\frac{\phi_0^2}{M_{\rm Pl}^2}.
\end{eqnarray}
Note that all the derivatives are of the same order in $\delta \xi$ and
$\frac{\phi_0^2}{M_{\rm Pl}^2}$, with higher-order runnings {\it increasing} in magnitude.
It is therefore not possible to model these spectra as a Taylor expansion in ${\rm
ln}k$, necessitating the use of Eq. (\ref{spec}) for imposing constraints.

In the case of perturbations generated through modulated reheating/curvatons, it is important to
mention that minimally coupled models have power spectra proportional
to the Hubble parameter formulated in the Einstein frame,  $P(k) \propto H^2_{\rm E}$.  In
contrast, non-minimally coupled models have spectra related to the Jordan frame Hubble parameter,
$P(k) \propto H^2_{\rm J}$.  While this distinction is not very important when determining the
duration of inflation ($H_{\rm E}$ and $H_{\rm J}$ differ by a factor $\phi^2/M_{\rm Pl}^2$), it is
of importance for the observables, which are themselves of order $\phi^2/M_{\rm Pl}^2$.
\footnote{In the absence of matter fields, there is no physical 
differences between minimally coupled and non-minimally coupled models. 
However, in the presence of matter fields, we should decide the frame in 
which matter fields are introduced. In the present paper, by 
non-minimally (or minimally) coupled models, we mean that light fields 
responsible for modulated reheating/curvaton are coupled minimally to 
the frame in which inflaton has non-minimal (or minimal, respectively) 
coupling to gravity.}

\section{Cosmological Constraints}
We now obtain cosmological constraints on the power spectrum of curvature
perturbations derived in the last section, Eq. (\ref{spec}).
In order to ensure that the inflaton perturbations are suppressed on CMB scales, and for
additional reasons that will become apparent in Section V, we consider low-scale inflation. We
therefore do not include a gravitational
wave contribution to the temperature anisotropy.  We utilize Markov Chain
Monte Carlo to obtain Bayesian parameter constraints
from the 5$^{th}$-year WMAP cosmic microwave background data
\cite{Hinshaw:2008kr,Gold:2008kp,Nolta:2008ih,Dunkley:2008ie,Komatsu:2008hk} and the
4$^{th}$ release SDSS
Luminous Red Galaxy (LRG) galaxy power spectrum data \cite{AdelmanMcCarthy:2005se}.  We use the
\texttt{CosmoMC} \cite{Lewis:2002ah} code with a modified version of \texttt{CAMB} \cite{Lewis:1999bs} in order
to compute the $C_\ell$-spectra arising from Eq. (\ref{spec}).
MCMC techniques \cite{Christensen:2000ji,Christensen:2001gj,Verde:2003ey,Mackay}
generate samples from the likelihood of model parameters,
\(\mathcal{L}({\bf d}|{\bm \theta})\),
where \({\bf d}\) represents the n-dimensional data and \({\bm \theta}\) the n-dimensional parameter
vector.
Via Bayes Theorem, the likelihood function
relates any prior knowledge about the parameter values, \(\pi({\bm \theta})\), to the posterior
probability distribution,
\begin{equation}
\label{bayes}
p({\bm \theta}|{\bf d}) = \frac{\mathcal{L}({\bf d}|{\bm \theta})\pi({\bm \theta})}{P({\bf d})} =
\frac{\mathcal{L}({\bf d}|{\bm \theta})\pi({\bm \theta})}{\int
\mathcal{L}({\bf d}|{\bm \theta})\pi({\bm \theta})d{\bm \theta}}.
\end{equation}
The posterior probability distribution function of a single parameter \(\theta_i\) is
obtained by marginalizing \(p({\bm \theta}|{\bf d})\) over the remaining parameters,
\begin{equation}
p(\theta_i|{\bf d}) = \int p({\bm \theta}|{\bf d})d\theta_1 \cdots d\theta_{i-1} d\theta_{i+1}\cdots d\theta_{n-1}.
\end{equation}
We adopt a four parameter base cosmology: the baryon density $\Omega_b
h^2$, the cold dark matter density $\Omega_c h^2$, the ratio of the sound
horizon to the angular diameter
distance at decoupling, $\theta_s$, and the optical depth to
reionization, $\tau$.  The power spectrum Eq. (\ref{spec})
is constrained by
varying the parameters $P(k_0)$, $A$, and $B$ directly in the Markov
chains, at the pivot scale $k_0 = 0.002\,h{\rm Mpc}^{-1}$.  We
also marginalize over the contribution from the Sunyaev-Zeldovich effect by varying the amplitude
$A_{\rm SZ}$ \cite{Komatsu:2002wc,Komatsu:2008hk},
assume purely adiabatic initial fluctuations, impose spatial flatness, and adopt
a top-hat prior on the age of the universe: $t_0 \in [10,20]$ Gyrs.  

The posterior probability distribution in the directions of $A$ and $B$ is expected to be non-Gaussian.
This is because the lines $A = 0$ and $B = 0$ give exactly scale invariant spectra, providing good fits
to the data.  The posterior thus has a high likelihood spine across the full prior of $A$ along $B =
0$, and conversely across the full prior of $B$ along $A=0$.  These
spines are very narrow, since, for example,  as
one moves to larger values of $B$ along the $A=0$ line, a small step $\delta A$ gives a strongly
non-power law spectrum.  Therefore, while the posterior along these
directions has high mean likelihood, the marginalized likelihood tends to be small.  Sampling from
this distribution thus poses a challenge, and we therefore utilize multicanonical sampling
\cite{berg,Gubernatis} in lieu of the more
standard Metropolis-Hastings algorithm.  Multicanonical sampling is useful for probing into the tails of
non-Gaussian distributions.
We measure convergence across eight chains using the Gelman-Rubin R
statistic ($R < 1.1$). 

The power spectrum Eq. (\ref{spec}) is not a freely tunable function. 
One requirement on the form of the spectrum is that the spacetime must be inflationary while astrophysical scales
are generated.  Otherwise, causally generated quantum fluctuations would not be stretched to
super-horizon scales.
Defining the parameter $\epsilon =
-\dot{H}/H^2$, and with $P(k) \propto H^2$, we have
\begin{equation}
\epsilon = \frac{1}{2}\frac{d{\rm ln}P(k)}{d{\rm ln}k}\left[\frac{1}{2}\frac{d{\rm ln}P(k)}{d{\rm ln}k} - 1\right]^{-1},
\end{equation}
with
\begin{equation}
\frac{d{\rm ln}P(k)}{d {\rm ln}k} =
-AB\left(\frac{k}{k_0}\right)^{-B}\left[1+A\left(\frac{k}{k_0}\right)^{-B}\right]^{-1}.
\end{equation}
This is enforced by simply requiring that $\epsilon < 1$ across the range of scales probed by cosmological
observations, corresponding to wavenumbers $\Delta {\rm ln}k \approx 10$.
Combinations of $A$ and $B$ which
violate this condition are rejected from the sample by assigning such
models very low likelihood.

As a means of comparison with the conventional parameterization, we transform
the parameters $A$ and $B$ into the spectral parameters $n_s(k_0)$ and
$\alpha(k_0) = dn_s(k_0)/d{\rm ln}k$,
\begin{eqnarray}
\label{trans}
n_s(k_0)  &=& 1 - \frac{AB}{1+A},\\
\alpha(k_0) &=& \frac{AB^2}{(1+A)^2}.
\end{eqnarray}
While the prior $\pi(A,B)$ is chosen to be flat, the prior distribution $\pi(n_s,\alpha)$ that results from this
transformation is not.  The Jacobian of the transformation
\begin{equation}
\pi(A,B) = \left|\frac{\partial(A,B)}{\partial(n_s,\alpha)}\right| \pi(n_s,\alpha),
\end{equation}
is in fact zero at the point ($n_s = 1$, $\alpha = 0$), and so the prior is not well-defined at this point.  While
this presents a formal difficulty, it is of little consequence in practice since the Markov chains do not sample the
neighborhood of this point with sufficient precision for the results to be affected by the prior in this area.  We
have verified that the posterior distribution is dominated by the likelihood across the prior range of $n_s$ and
$\alpha$, and so $\pi(n_s,\alpha)$ is effectively flat in this region.  

Figure 1 displays the  1-$\sigma$ (68\% CL) and 2-$\sigma$ (95\% CL) error contours for the rapid roll model compared 
to the constraints obtained on $n_s$ and $\alpha$ when they are
allowed to vary as free parameters (hereafter referred to as the plr
model), as done in standard analyses by adopting
the form
\begin{equation}
\label{plr}
{\rm ln}\frac{P(k)}{P(k_0)} = (n_s - 1){\rm ln}\frac{k}{k_0} +
\frac{1}{2}\alpha {\rm ln}^2 \frac{k}{k_0}.
\end{equation}
\begin{figure}
\centerline{\includegraphics[width=3.0in]{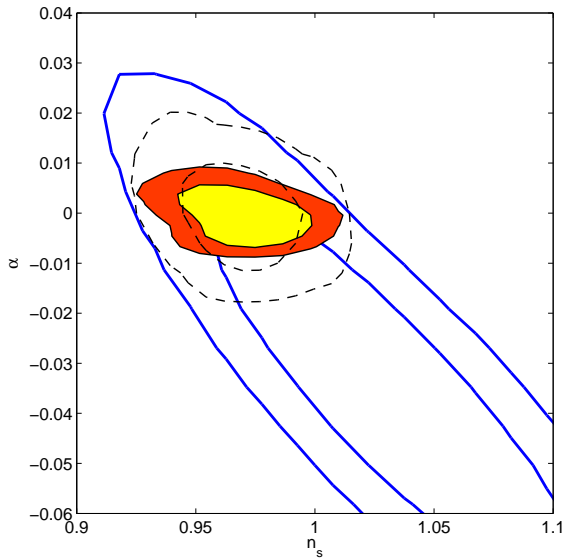}}
\caption{Marginalized constraints on rapid roll inflation as compared to constraints obtained on the spectral parameters in the
standard definition (plr) Eq. (\ref{plr}).  The blue contours denote 68\% and
95\% CL for the plr model with WMAP5+SDSS+ACBAR.  The filled
(orange-yellow) contours denote the same for the rapid roll model.  The
black dashed contours are constraints on rapid roll inflation with only
WMAP5+SDSS. We shall see in figure 4 that, in D-brane inflation, these
 constraints for the rapid roll model are consistent with wide range of 
 $|\dot{\phi}/H\phi|$.} 
\end{figure}
The rapid roll model is significantly more predictive than the
model-independent, plr spectrum with a preference for
$n_s < 1$ and small running.  The marginalized
constraints are given in Table 1.
\begin{table}
\begin{tabular}{||l|c|c|c||}
\hline
Model & $n_s(k_0)$ & $\alpha(k_0)$ & $-2{\rm ln}\hat{\mathcal{L}}$\\
\hline
rapid roll &\,$0.972^{+0.022}_{-0.031}$\, &\,
$-{5.01\times10^{-3}}^{+0.01}_{-0.001}$\,& \,2693.5\,\\
\hline
plr &\, $1.04^{+0.16}_{-0.21}$\, & \,$-0.035^{+0.024}_{-0.025}$\,&
\,2691.2\,\\
\hline
\end{tabular}
\caption{Marginalized 2-$\sigma$ errors on the spectral parameters for the
rapid roll model and the standard parameterization for WMAP5+SDSS+ACBAR.}
\end{table}
The tight constraints on the rapid roll parameters are a result of the
correlations between $n_s$, $\alpha$, and the infinite tower of
higher-order $k$-dependent terms (e.g. Eq. (\ref{runnings})).  The running of the
running, $\beta = d\alpha/d{\rm ln}k$, and all terms of higher-order, are
completely determined by $n_s$ and $\alpha$.  For example,
\begin{equation}
\beta = \frac{\alpha^2}{n_s-1},
\end{equation}
and so as the running is increased (for $n_s \neq 1$), $\beta$ likewise
increases in magnitude.  The tight constraints on $\alpha$ are therefore
a direct result of the data's intolerance to a large $\beta$.  
The constraints on the upper (lower) value
of $n_s$ are a result of the data's requirement that blue spectra be
accompanied by significant negative (positive) running, as indicated by the black
contours in Figure 1.  The magnitude of running permitted in the rapid roll
model is insufficient to accommodate very tilted spectra. 

Of the spectra lying within the 2-$\sigma$ envelope, the most distinctive
are those that exhibit a relatively abrupt increase in power
on small scales at around $0.02\,h{\rm Mpc}^{-1}$.  This $k$-dependence
is a result of higher-order runnings.  Since the WMAP data becomes noise-dominated at around the second
peak, the addition
of small-scale CMB measurements might have an effect on
constraints.  In Figure 2 we present the $C_\ell^{TT}$-spectrum of the
best fit plr model (solid line) along with a model lying on
the outskirts of the 2-$\sigma$ envelope (dashed line) when only WMAP5+SDSS are used. As suggested by Figure 2 and confirmed in Figure 1, the inclusion of the latest ACBAR data-set 
\cite{Reichardt:2008ay} 
yields a distinctive improvement over the WMAP+SDSS only constraints,
ruling-out such spectra at greater than 95\% CL. 
\begin{figure}
\centerline{\includegraphics[angle=270,width=3.5in]{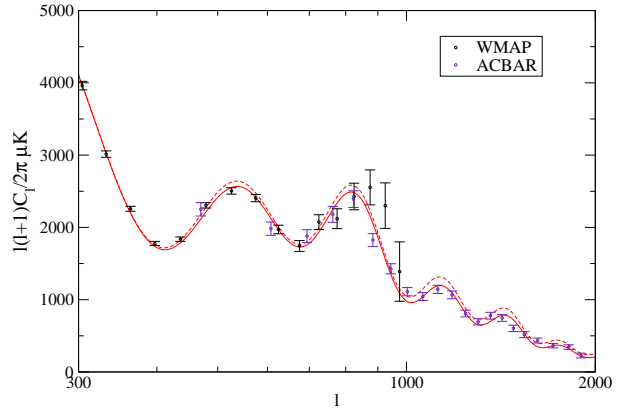}}
\caption{Best-fit plr model (red solid) as compared to a rapid roll spectrum lying at the edge of the
95\% CL (red dashed).  WMAP 5-year data (black points) and the 2008 ACBAR data (purple points) are
included.  The inclusion of ACBAR data imposes additional constraints on rapid roll models which exhibit 
higher-order running on small scales.}
\end{figure}

We also note that the rapid roll model does a poorer job of fitting the data
than the plr model, with a difference in
effective chi-square, $\Delta \chi^2 = -2\Delta {\rm ln}\mathcal{L}
\approx 2$, between them.  Of course, this is not to say that the
plr model is necessarily preferred by the data, since a proper comparison must also incorporate the number of degrees of
freedom available to each model.  However, this is not as simple as just counting the tunable parameters (both models have 8 --
with 3 defining the power spectrum), since this number is not representative of {\it model
complexity}, the quantity of interest when performing Bayesian model selection \cite{Kunz:2006mc}.  Rather, one
should identify the number of parameters that are sufficiently constrained by the data, giving an
{\it effective} number of parameters, $\mathcal{C}$ \cite{Kunz:2006mc,Liddle:2007fy},
\begin{equation}
\mathcal{C} = \overline{\chi^2(\bm{\theta})} - \chi^2(\hat{\bm{\theta}}),
\end{equation}
where $\bm{\theta}$, is the parameter vector, the bar indicates an average, and the hat denotes the best-fit value.   Using this
prescription, we find that the plr model has 7 effective degrees of freedom, while
the rapid roll model
has 6.7.\footnote{The Sunyaev-Zeldovich amplitude $A_{\rm SZ}$ is poorly constrained by the data and
therefore does not contribute to this figure.}  The fact that the rapid roll model has fewer effective
parameters can be understood by examining the ansatz Eq. (\ref{spec}).  As mentioned, both $A$ and $B$
give high mean likelihood across the full range of their respective priors.  While the marginalized
distribution of these parameters is well localized, there exist directions in the prior 
volume along which these parameters are unconstrained.  The effective complexity criterion therefore does
not count $A$ and $B$ as two free parameters, but rather a little less than this.
It is therefore appropriate to conclude that current data does not single-out a
preferred model, and it will be necessary for future experiments to improve our understanding of the
power spectrum.

\section{Rapid Roll D-Brane Inflation}

As a simple example of minimally coupled rapid-roll inflation (see also
the appendix), let us
consider D3/$\overline{\mathrm{D}3}$ brane inflation in a warped 
throat~\cite{Kachru:2003sx}. In this model, inflation is driven while a
D3-brane is attracted towards an $\overline{\mathrm{D}3}$ sitting at the
throat tip. The radial position~$r$ of the D3 plays the role of the
inflaton through $\phi \equiv \sqrt{T_3} r$, and the warped tension of
the $\overline{\mathrm{D}3}$ sources inflation. It is known that the
inflaton receives a large mass from moduli stabilization effects and
that
this tends to prevent slow-roll. Here, we also take into account corrections due to
the throat coupling to the bulk Calabi-Yau space, which can be analysed
using gauge/gravity duality~\cite{Baumann:2008kq} (see also
\cite{Ceresole:1999zs,Ceresole:1999ht,Aharony:2005ez})\footnote{There
can also be additional contributions from moduli stabilization, as was
investigated in
\cite{Baumann:2006th,Baumann:2007np,Krause:2007jk,Baumann:2007ah} in
order to fine-tune the inflaton potential.}. We consider the case where
the leading bulk effect arises from a non-chiral operator of dimension
$\Delta=2$ in the dual CFT, where the inflaton potential takes
the form
\begin{equation}
 V(\phi) = 2 h_0^4 T_3 \left\{ 1- \frac{3 h_0^4 T_3}{8 \pi^2 \phi^4}
			+\mu 
 \left( \frac{\phi}{M_{\mathrm{Pl}}}\right)^2\right\},
 \label{eq-pot}
\end{equation}
where $h_0$ is the warping at the tip of the throat, $T_3$ is the brane tension, and the second term
denotes the D3-$\overline{\mathrm{D}3}$ Coulombic interaction. When the
bulk effects are absent, $\mu = 1/3$ as was derived in the original
KKLMMT paper~\cite{Kachru:2003sx}. The bulk effect introduces a negative
contribution to the inflaton mass-squared, i.e. makes $\mu$ smaller
than 1/3. 

We restrict ourselves to the region~$\phi^2 \ll M_{\mathrm{Pl}}^2$. When
the Coulombic attraction is negligible compared to the mass term:
\begin{equation}
 \frac{M_{\mathrm{Pl}}^2 h_0^4 T_3}{\mu \phi^6} \ll 1, \label{eq-neg}
\end{equation}
we obtain rapid-roll inflation with 
\begin{equation}
 3 M_{\mathrm{Pl}}^2 H_{\mathrm{inf}}^2 \simeq 2 h_0^4 T_3.
\end{equation}
From Eqs. (\ref{Eps}) and (\ref{Eta})  one can estimate
\begin{equation}
 \epsilon \simeq 2 \mu^2 \frac{\phi^2}{M_{\mathrm{Pl}}^2} ,\quad 
 \eta \simeq 2 \mu. 
\end{equation}
Then, setting
\begin{equation}
 c = \frac{3+\sqrt{9-24 \mu}}{2},
\end{equation}
the inflationary attractor Eq. (\ref{gfr}) gives
\begin{equation}
 \frac{H}{\dot{\phi}} \simeq -\frac{3+\sqrt{9-24 \mu}}{12 \mu
  }\frac{1}{\phi}. \label{eqn:dotphi-mu}
\end{equation}
We note that the coefficient of $1/\phi$ agrees with $-1/\mu_0$ from Eq. (\ref{ab}) by setting $\xi =
0$ and $v_2 = 2\mu$.
Upon estimating the number of efoldings that can be obtained, one should
note that the inflaton field range is bounded by the length of the
throat~\cite{Baumann:2006cd}. Specifically, for throats supported by $N$
units of D3-brane charge, the bound gives $\phi_{\mathrm{UV}} \sim
\sqrt{T_3} R$, where $R^4 \sim g_s N
\alpha'^2$~\cite{Gubser:1998vd}. Also, the end of inflation can be
identified with the time when Eq. (\ref{eq-neg}) breaks down, which gives
\begin{equation}
 \phi_{\mathrm{end}} \sim \left(\frac{M_{\mathrm{Pl}}^2 h_0^4
			   T_3}{\mu}\right)^{1/6}.
\end{equation}
Hence the maximum number of efoldings that can be obtained is
\begin{multline}
 N_{\mathrm{max}} = \int^{\phi_{\mathrm{end}}}_{\phi_{\mathrm{UV}}} H
  \frac{d\phi}{\dot{\phi}} \\
 \simeq 
 \frac{3+\sqrt{9-24 \mu}}{12 \mu }\ln
 \left( \frac{\mu^{1/6}N^{1/4} g_s^{1/4} \alpha'^{1/2}}{h_0^{2/3} L}
 \right).  \label{eq-maxN}
\end{multline}
Here we have used $T_3 \sim 1/(g_s \alpha'^2)$ and 
$M_{\mathrm{Pl}} \sim L^3/(g_s \alpha'^2)$ where $L$ is the typical
length scale of the six-dimensional bulk. The efolding
number Eq. (\ref{eq-maxN}) depends sensitively on the mass
parameter~$\mu$, hence, one sees that even a slight effect from the
bulk can change the duration of inflation drastically.
 
On the other hand, inflation should have started before the present
Hubble scale exited the horizon. Therefore the required number of
efoldings is at least
\begin{equation}
 N \simeq \ln \left( 10^{29}
  \frac{V_{\mathrm{inf}}^{1/4}}{M_{\mathrm{Pl}}}\right)
 \simeq \ln \left(10^{29}\frac{ h_0 g_s^{3/4} \alpha'^{3/2}}{L^3}\right),  
 \label{eq-reqN}
\end{equation}
where we have assumed instantaneous (p)reheating, i.e. the universe was
dominated by radiation right after inflation ended until
matter-radiation equality. (This is realized in the
case where there are D-branes left after inflation, since closed strings
generated from D3-$\overline{\mathrm{D}3}$ annihilation soon decay into
lighter states such as gravitons, KK 
modes, and open string modes on the residual D-branes, and then
thermalize. See e.g. \cite{Chen:2006ni}.) 

In Figure 3, we present the ratio between the field range required to obtain sufficient
inflation and the allowed range in the $\mu$ - $\log_{10}h_0$
plane. For example, if we take $g_s = 0.1$, $N = 10^3$, $L/\alpha'^{1/2}
= 10$, and $h_0 = 10^{-11}$ (which set the inflation energy
scale~$V_{\mathrm{inf}}^{1/4}$ and the local string
scale~$h_0/\alpha'^{1/2}$ to be of order 1 TeV, and $M_{\mathrm{Pl}}
\alpha'^{1/2} \sim 10^4$), then the necessary number of efoldings can be
obtained when $\mu \lesssim 0.2$ is satisfied. As can be seen from
Eqs. (\ref{eq-maxN}) and (\ref{eq-reqN}), smaller $h_0$ and $\mu$ are
preferred for enough inflation to occur. Let us emphasize that the
original Hubble scale mass (i.e. $\mu \sim 0.3$) need not be exactly
cancelled for rapid roll inflation, provided there is sufficient warping of the
throat.

\begin{figure}
\centerline{\includegraphics[width=3.0in]{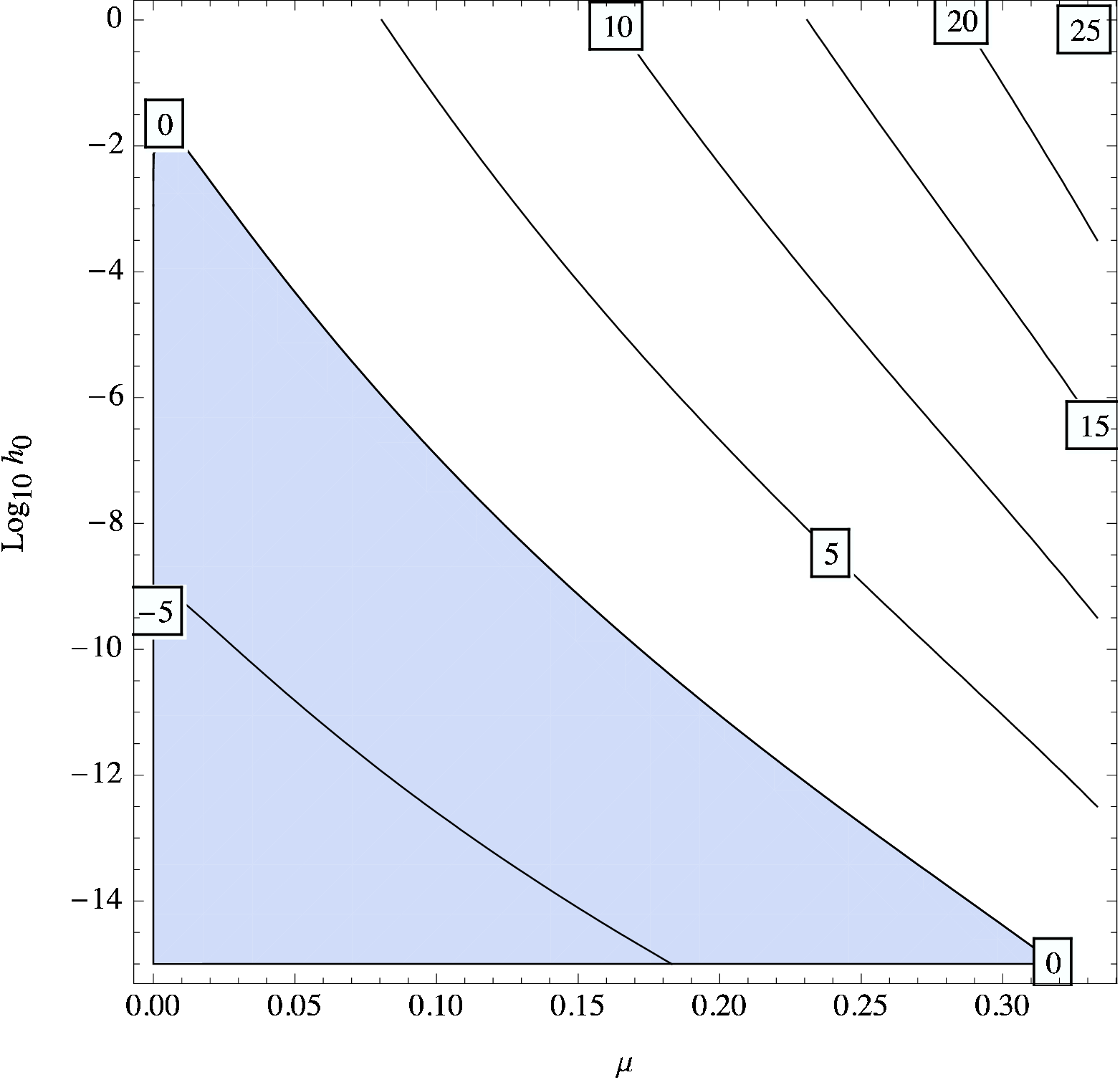}}
\caption{Contour plot of $\log_{10}(\phi_H / \phi_{UV})$, where $\phi_H$
 is the inflaton field value when the present Hubble scale exited the
 horizon. $\phi_H$ is estimated by assuming that the inflaton
 potential Eq. (\ref{eq-pot}) is valid throughout. The horizontal and
 vertical axes denote, respectively, $\mu$ and $\log_{10}
 h_0$. The other parameters are set to $g_s = 0.1$, $N=1000$, and
 $L/\alpha'^{1/2}  = 10 $. Necessary number of efoldings can be obtained
 in the shaded region.}
\end{figure}

The inflaton field value when the pivot scale exited the horizon can
similarly be estimated from Eqs. (\ref{eq-maxN}) and (\ref{eq-reqN}), 
\begin{equation}
 \frac{\phi_0}{M_{\mathrm{Pl}}} \sim \left(\frac{V_{\mathrm{inf}}}{\mu
				      M_{\mathrm{Pl}}^4}\right)^{1/6} 
 \left(10^{28}
  \frac{V_{\mathrm{inf}}^{1/4}}{M_{\mathrm{Pl}}}\right)^{\frac{12\mu}{3+\sqrt{9-24
 \mu}}}. \label{eq-pivot}
\end{equation}
Note that for the pivot scale, the term inside the log in
Eq. (\ref{eq-reqN}) differs by an order of magnitude.

The spectral index and its running of the Hubble-squared during
inflation can be estimated from Eq. (\ref{spec}) with $\xi=0$ and
$v_2=2\mu$, or from Eqs. (\ref{eq-A20}) and (\ref{eq-A21}), 
\begin{equation}
 n_s-1 \simeq -\frac{18}{c^2}\cdot 2
  \mu^2\frac{\phi^2}{M_{\mathrm{Pl}}^2}, \label{eq-ns}
\end{equation}
\begin{equation}
 \frac{d n_s}{d \ln k} \simeq \frac{36}{c^2}(3-c)\cdot 2 \mu^2
  \frac{\phi^2}{M_{\mathrm{Pl}}^2},  \label{eq-running}
\end{equation}
where the right hand sides are given at the moment of horizon
crossing Eq. (\ref{eq-pivot}). One clearly sees a linear relation between
Eqs. (\ref{eq-ns}) and (\ref{eq-running}), 
\begin{equation}
\label{pred}
 \frac{dn_s}{d\ln k} \simeq -\left(3-\sqrt{9-24\mu}\right) (n_s-1).
\end{equation}

In Figure 4, we compare this prediction to the constraints obtained in
Section IV.  We plot the prediction Eq. (\ref{pred}) for the values 
$\mu = 0.001$, $\mu = 0.05$, and $\mu = 0.3$ to illustrate the range of
values of $\alpha$ and $n_s$ consistent with current data.  These values
correspond to $|\dot{\phi}/H\phi|\simeq 0.002$,
$|\dot{\phi}/H\phi|\simeq 0.1$, $|\dot{\phi}/H\phi|\simeq 0.8$,
respectively. (See (\ref{eqn:dotphi-mu}).) For fixed $\mu$, each
prediction is a function of $\phi_0$ which depends on the inflationary
energy scale. The rapid roll brane inflation model is consistent with
current data for a wide range of $\mu$ and, thus, $\dot{\phi}/H\phi$ and
$\eta$. 
\begin{figure}
\centerline{\includegraphics[width=3.0in]{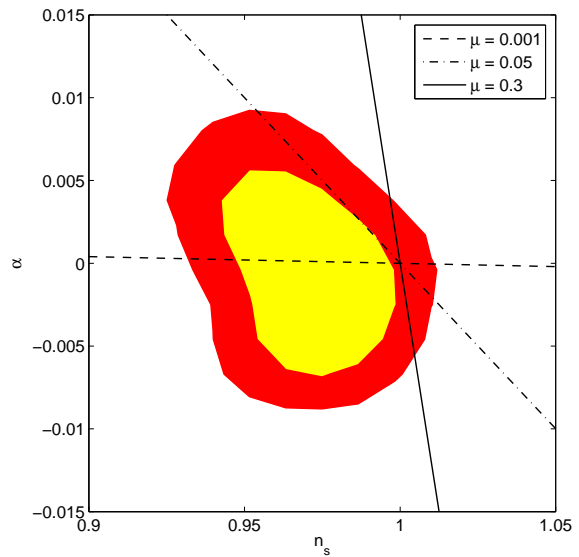}}
\caption{Prediction of the rapid roll brane model Eq. (\ref{pred}) for the values $\mu =
0.001$, $\mu = 0.05$,  and $\mu = 0.3$ compared to the constraints
obtained in Section IV. These values correspond to
$|\dot{\phi}/H\phi|\simeq 0.002$, $|\dot{\phi}/H\phi|\simeq 0.1$, 
$|\dot{\phi}/H\phi|\simeq 0.8$, respectively.} 
\end{figure}

Finally, we mention that even though we have treated D-brane inflation
as a minimally coupled model with Hubble scale mass, interpreting the
mass as arising from the inflaton's conformal coupling to gravity does
not make much difference for the inflaton dynamics and the efolding
number that can be obtained. However, it does make distinct differences
for the spectral index and its running of the Hubble-squared, for reasons discussed at the end of
Section III. One can
explicitly see this for the D-brane inflation case by comparing the
above results with that of \cite{Kobayashi:2008rx}.

\section{Conclusions}
We have obtained cosmological constraints on rapid roll models of inflation.  Rapid
roll solutions exist as attractors for tree-level hybrid-type potentials with a range
of mass terms, with both minimal and nonminimal gravitational couplings.  Such solutions are therefore relevant to model building in string
theory.  We considered perturbations generated through modulated reheating and/or the
curvaton scenario instead of the observationally unacceptable inflaton-generated
perturbations in these models.  We obtained an analytic expression for the power
spectrum in this case, 
\begin{equation}
P(k) = \frac{P(k_0)}{1+A}\left[1 +
A\left(\frac{k}{k_0}\right)^{-B}\right],
\end{equation}
observing a lack of a hierarchy amongst lower-order and
higher-order $k$-dependencies.  We have performed a Bayesian analysis on this power spectrum
using the latest CMB and LSS data.  The higher-order runnings are fully determined by
$n_s$ and $\alpha$, and so these parameters are tightly constrained relative to
general
power-law + running spectra.  We find that rapid roll models are constrained to have
$n_s < 1$ and small running, $|\alpha| < 0.01$.  These tight predictions make
possible the falsifiability of rapid roll inflation with upcoming CMB missions.  The
higher-order runnings that occur in these spectra might also be further constrained
via upcoming astrophysical probes, such as the 21-cm line of neutral hydrogen clouds.  

We also construct a realistic model of rapid roll brane inflation.  We find that it
is possible to obtain sufficient inflation even in the presence of the large mass
terms that arise from moduli stabilization in these models.  The power spectrum
generated by this model is in good agreement with the constraints obtained in this
work, across a range of mass scales.   Therefore, while the inflaton-generated
perturbations of such models are not observationally viable, the spectra generated
through other means, such as modulated reheating or curvatons, breathe new life into
these constructions.  Additionally, the tight constraints imposed on these spectra
suggest that future CMB data has the potential to further restrict them, or rule such
models out entirely.

\section{acknowledgements}
The calculations were performed by the EUP
cluster system installed at Graduate School of
Frontier Sciences, University of Tokyo.
We thank Damien Easson for comments on a draft version of this paper.
The work of S.M. was supported in part by MEXT through a Grant-in-Aid 
for Young Scientists (B) No.~17740134, and by JSPS through a 
Grant-in-Aid for Creative Scientific Research No.~19GS0219.
This work was supported by World Premier International
Research Center Initiative (WPI Initiative), MEXT, Japan.

\appendix

\section{Discussions on Minimally Coupled Rapid-Roll Inflation}

We investigate the dynamics of a rapid-rolling inflaton which minimally
couples to gravity. The calculations carried out in this appendix is
analogous to that of \cite{Kobayashi:2008rx} where a conformally
coupled inflaton was studied.

\subsection{Conditions for Rapid-Roll Inflation}

We consider a minimally coupled inflaton with the Lagrangian
\begin{equation}
 \mathcal{L} = \sqrt{-g} \left[
\frac{M_{\mathrm{Pl}}^2}{2} R - \frac{1}{2} g^{\mu\nu} \partial_{\mu}\phi
\partial_{\nu}\phi - V(\phi) \right].
\end{equation}
Choosing a flat FRW background, we obtain the Friedmann equation
\begin{equation}
3  M_{\mathrm{Pl}}^2 H^2 = \frac{1}{2}\dot{\phi}^2 + V(\phi),
\label{eq-1}
\end{equation}
and the equation of motion of $\phi$,
\begin{equation}
 \ddot{\phi} + 3 H \dot{\phi} + V'(\phi) = 0 .
\label{eq-2}
\end{equation}
We claim that during inflation, the inflaton dynamics is well described
by the following approximations
\begin{equation}
 3 M_{\mathrm{Pl}}^2 H^2 \simeq V, \label{eq-3}
\end{equation}
\begin{equation}
 c H \dot{\phi} \simeq -V', \label{eq-4}
\end{equation}
where $c$ is a dimensionless constant. 

Let us define the following parameters,
\begin{equation}
 \epsilon \equiv \frac{M_{\mathrm{Pl}}^2}{2}
  \left(\frac{V'}{V}\right)^2, \quad
 \bar{\eta} \equiv \eta + \frac{c^2}{3} - c,
\end{equation}
where 
\begin{equation}
 \eta \equiv M_{\mathrm{Pl}}^2 \frac{V''}{V}.
\end{equation}
Then the necessary conditions for the approximate relations~(\ref{eq-3})
and (\ref{eq-4}) to hold can be derived, respectively, as
\begin{equation}
 \frac{3}{c^2}\epsilon \ll 1, \quad
 \frac{3}{c^2} \left| 3 \epsilon - \bar{\eta} \right| \ll 1. \label{eq-nc}  
\end{equation}
Note that we have taken a time-derivative of both sides of the
approximate relation~(\ref{eq-4}) and then substituted it into
(\ref{eq-2}) in order to derive the necessary condition for
(\ref{eq-4}). When $c \sim \mathcal{O}(1)$, the necessary
conditions~(\ref{eq-nc}) reduce to simple forms
\begin{equation}
 \epsilon \ll 1,\quad \left| \bar{\eta} \right| \ll 1.
\end{equation}
This shows that the parameters which should be minimized in order to
realize inflation are $\epsilon$ and $\bar{\eta}$ (instead of $\epsilon$ and
$\eta$). The constant $c$ in (\ref{eq-4}) is the {\it largest} constant
that minimizes $\bar{\eta}$, as we will show in the next section.

\subsection{Stability of the Attractor}

Let us define $\beta$ to parametrize the validness of the attractor
(\ref{eq-4}) as
\begin{equation}
 c H \dot{\phi} = -V' (1+\beta). \label{eq-beta}
\end{equation}
We derive an evolution equation of $\beta$ in this section.

Some useful relations are
\begin{multline}
 \Sigma \equiv \frac{3 M_{\mathrm{Pl}}^2 H^2}{V}   \\
 = \frac{1}{2} \left\{1+ \sqrt{1+\frac{12
 (1+\beta)^2}{c^2}\epsilon}\right\} = 
 1+\mathcal{O(\epsilon)}, 
\end{multline}
\begin{equation}
 \frac{\dot{\phi}^2}{V } = 2 \left(\Sigma-1\right), \quad
 \frac{\dot{H}}{H^2} = - \frac{3 (\Sigma-1)}{\Sigma}.
\end{equation}
Then by time-differentiating both sides of (\ref{eq-beta}),
one obtains 
\begin{multline}
 \frac{\dot{\beta}}{H} 
 = \frac{3}{\Sigma} 
 \left\{ 1-2\Sigma + \frac{1+\beta}{c} \eta\right\} (1+\beta) + c \\
 = \beta (3-2 c) + \beta^2 (3-c) + \mathcal{O}(\epsilon,\, \bar{\eta}).
\label{eq-star}
\end{multline}
Once $\beta$ becomes small, the far right hand side is dominated by the
linear term (i.e. $\beta (3-2c)$). One can see that as long as
\begin{equation}
 3-2c<0,  \label{eq-cond}
\end{equation}
$\beta$ damps as the universe expands, and eventually its value settles
down to that corresponding to the source terms,
\begin{equation}
 \beta = \mathcal{O}(\epsilon,\, \bar{\eta}). \label{eq-damp}
\end{equation}
Hence the the condition (\ref{eq-cond}) is required for the relations
(\ref{eq-3}) and (\ref{eq-4}) to be an inflationary attractor. 

\vspace{\baselineskip}
Next we study the values of $c$ chosen for inflation. As we stressed in
the previous section, $c$ is a constant which minimizes $\bar{\eta}/c^2$. 

\subsubsection*{Case Study: negligible $\eta$}

This corresponds to the familiar slow-roll inflation. Here $\bar{\eta}
/c^2 = 0$ gives $c=3$, which realizes the slow-roll approximations. It
is clear that the stability condition~(\ref{eq-cond}) is satisfied. 

\subsubsection*{Case Study: non-negligible constant $\eta$}

$\bar{\eta} /c^2 =0$ is solved by
\begin{equation}
 c = \frac{3\pm\sqrt{9-12\eta}}{2} \equiv c_{\pm}. 
\end{equation}
Since $3-2c_{\pm} = \mp \sqrt{9-12 \eta} \lessgtr 0$, one sees from
(\ref{eq-cond}) that the larger solution $c_+$ is chosen for the
inflationary attractor,
\begin{equation}
 c = \frac{3 + \sqrt{9-12\eta}}{2}. \label{cplus}
\end{equation}
It should also be noted that $\eta \leq 3/4$ needs to be satisfied.

\subsection{Scale Dependence of $H^2$}

By using the relations in the previous section and considering $\beta$
to be damped to (\ref{eq-damp}), one can compute the time
differentiations of the Hubble parameter,
\begin{equation}
 \frac{\dot{H}}{H^2} = -\frac{9}{c^2} \epsilon +
  \mathcal{O}(\bar{\eta}\epsilon,\, \epsilon^2),
\end{equation}
\begin{equation}
 \frac{1}{H}\left( \frac{\dot{H}}{H^2}\right)^{\cdot} = -\frac{18}{c^2}
  (c-3) \epsilon + \mathcal{O}(\bar{\eta} \epsilon,\, \epsilon^2).
\end{equation}
Especially, the spectral index and its running of the Hubble-squared are
($k=aH$), 
\begin{multline}
 n_s - 1 = \frac{d\ln H^2}{d \ln k}= 
 2  \frac{\dot{H}}{H^2} \left(1+\frac{\dot{H}}{H^2}\right)^{-1}  \\
 = -\frac{18}{c^2}\epsilon + \mathcal{O}(\bar{\eta} \epsilon,\, \epsilon^2),
 \label{eq-A20}
\end{multline}
\begin{multline}
 \frac{d n_s}{d \ln k} = 
2\left(1+\frac{\dot{H}}{H^2}\right)^{-3}\frac{1}{H}
\left( \frac{\dot{H}}{H^2}\right)^{\cdot} \\
 = \frac{36}{c^2}(3-c) \epsilon + \mathcal{O}(\bar{\eta} \epsilon,\,
\epsilon^2) . 
 \label{eq-A21}
\end{multline}
For inflaton potentials of the form (\ref{v}), these expressions
match with what follows from Eq. (\ref{spec}) with $\xi = 0$.

\end{document}